\documentclass{article}
\usepackage[T1]{fontenc} 
\usepackage[utf8]{inputenc} 
\usepackage{ismir,amsmath,cite,url}
\usepackage{graphicx}
\usepackage{color}
\usepackage{amssymb}
\usepackage{booktabs}
\usepackage{multirow}

\graphicspath{ {./figs/} } 

\usepackage{lineno}

\title{Equivariant self-supervision for musical tempo estimation}

\oneauthor
 {Elio Quinton}
 {Universal Music Group}

\def\authorname{E. Quinton}

\begin{document}

\maketitle
\begin{abstract}

Self-supervised methods have emerged as a promising avenue for representation learning in the recent years since they alleviate the need for labeled datasets, which are scarce and expensive to acquire. 
Contrastive methods are a popular choice for self-supervision in the audio domain, and typically provide a learning signal by forcing the model to be invariant to some transformations of the input. These methods, however, require measures such as negative sampling or some form of regularisation to be taken to prevent the model from collapsing on trivial solutions. 
In this work, instead of invariance, we propose to use equivariance as a self-supervision signal to learn audio tempo representations from unlabelled data. We derive a simple loss function that prevents the network from collapsing on a trivial solution during training, without requiring any form of regularisation or negative sampling.
Our experiments show that it is possible to learn meaningful representations for tempo estimation by solely relying on equivariant self-supervision, achieving performance comparable with supervised methods on several benchmarks. 
As an added benefit, our method only requires moderate compute resources and therefore remains accessible to a wide research community.

\end{abstract}
\section{Introduction}
\label{sec:intro}

Manually annotated data is scarce and expensive to acquire, becoming a bottleneck to the continued progress of supervised learning methods in recent years. This is particularly true in the music domain, where a large proportion of content available is under copyright. 
Unlabelled data is comparatively abundant and inexpensive. Self-Supervised Learning (SSL) methods, which do not require labeled training data, have shown very promising development by rendering larger unlabelled datasets usable for training and yielding performance comparable or surpassing supervised benchmarks on Natural Language Processing (NLP) \cite{raffel_exploring_2019, devlin_bert:_2018}, computer vision tasks \cite{chen_simple_2020}, speech processing \cite{schneider_wav2vec_2019} and music tasks \cite{spijkervet_contrastive_2021}.
Self-supervised learning is still only nascent in musical audio, in comparison with other domains. To date it has only been considered with target downstream tasks that are a limited subset of the Music Information Retrieval (MIR) field, e.g. auto-tagging, genre classification or cover song detection \cite{spijkervet_contrastive_2021, yao_contrastive_2022, zhao_s3t_2022}. 
With this contribution, we propose to extend the application of a SSL framework to the rhythmic properties of music by applying it to the tempo estimation task, which state of the art is still within the realm of supervised methods \cite{bock_accurate_2015, schreiber_single-step_2018, bock_deconstruct_2020, sun_musical_2021}.

Self-supervision is typically realised by creating a pretext task providing a training signal from which it is expected that the model can learn useful representations. The general intuition underpinning this family of approaches is that accurately performing on the pretext task requires the model to learn meaningful representations of the domain at hand. 
For domains where data is made of discrete tokens, such as text, pretext tasks like masked language modelling or next sentence prediction as have proven to be effective and brought a step change in NLP \cite{devlin_bert:_2018}. Taking inspiration from it, comparable approaches can be devised in the symbolic music domain, which data is also in the form of discrete tokens \cite{huang_music_2018-1}.  
In domains where data is dense and continuous, such as images or audio, one option is to apply similar token-based methods to discrete representations of the dense input \cite{baevski_vq-wav2vec_2019, dhariwal_jukebox_2020}. 
Alternatively, Siamese network frameworks \cite{hadsell_dimensionality_2006}, where two models process two views of a given input, are a popular choice for handling continuous data directly. In particular, contrastive methods where the pretext task consists in discriminating whether two inputs (or set thereof) should yield a similar, or dissimilar representation have been shown to be effective in computer vision \cite{chen_simple_2020} and in the audio  domain \cite{jiang_speech_2020, wang_towards_2022, manocha_cdpam_2021, srivastava_conformer-based_2022}, including musical audio \cite{spijkervet_contrastive_2021, yao_contrastive_2022, zhao_s3t_2022}. 

In the contrastive learning framework, for every training sample, two views are generated by applying random data augmentations. 
The training objective then constrains the model to produce representations that are \emph{invariant} to the augmentations applied to the training data and yet discriminative between different samples. 
The choice of augmentations is a critical design choice that impacts the quality and properties of representation learnt \cite{tian_what_2020}. 
For example, in \cite{spijkervet_contrastive_2021} the model is trained to be invariant to pitch-shifting. While this is appropriate for fine-tuning on an auto-tagging task, it would not be suitable for downstream tasks such as chord or key estimation.
Generally speaking, the invariance constraints explicitly specifies what the representations should \emph{not} be sensitive to.
Conversely, \emph{equivariant} constraints can be applied to enforce preferences on what features or concepts may be desirable to capture in the learnt representations \cite{dangovski_equivariant_2021, wang_self-supervised_2020}.
This is particularly attractive in scenarios where it is not clear how invariance constraints can yield a suitable representation, such as F0 estimation \cite{gfeller_spice_2020}, or for guaranteeing that the representations capture specific semantically meaningful dimensions such as harmony or rhythm, which may be beneficial for applications such as music search and discovery. 

In this work, instead of \emph{invariance}, we propose to use \emph{equivariance} as a self-supervision constraint to learn audio representations that specifically capture musical tempo, from unlabelled data. 
We propose a Siamese network framework where we produce two views of each training sample by applying a time stretching transformation with random factor and impose an equivariance constraint between the representations. 
Although the tempo of the training samples is unknown, as a result of the time-stretching transformation, the tempo of each sample is modified in two different, but known, ways. 
From there we derive a loss function that exploits this information and enforces the equivariant constraint. 
Because we wish the representations to capture rhythmic properties of music only, we also add audio augmentations to promote robustness against potentially confounding attributes of the audio signal.
As is customary with self-supervised pre-training \cite{chen_simple_2020, spijkervet_contrastive_2021}, we evaluate the quality of the representations learnt with our method by freezing the model and fine tuning a linear layer on a downstream tempo estimation task.

Our key contributions are the following. 
1- To the best of our knowledge, our work is the first to rely solely on an equivariance-based objective to learn musical audio representations.
2- We derive a simple loss function that formally prevents the network from collapsing on a trivial constant solution during training, without requiring any form of regularisation or negative sampling.
3- Our experiments show that it is possible to learn meaningful representations for tempo estimation by solely relying on equivariant self-supervision, and to achieve performance comparable with supervised methods on several benchmarks.
4- Our experiments demonstrate out-of-domain transferability of the representations learnt across multiple datasets.
5- As an added benefit, our method only requires moderate compute resources and therefore remains accessible to a wide research community. In order to further reproducibility we make code and pre-trained models available\footnote{https://github.com/Quint-e/equivariant-self-supervision-tempo}.

\section{Background}
\label{sec:related_work}

\subsection{Tempo Estimation}

In many musical traditions, tempo is one of the fundamental characteristics of music. Tempo estimation was also one of the first tasks to have been explored in the field of MIR \cite{scheirer_tempo_1998, cemgil_tempo_2000, dixon_automatic_2001, alonso_study_2003, gouyon_experimental_2006}.
Historically, tempo estimation was performed by first extracting an onset detection function \cite{bello_tutorial_2005, bock_maximum_2013} and then perform tempo estimation via the computation of inter-onset intervals or some form of periodicity function from the onset detection curve \cite{scheirer_tempo_1998, dixon_automatic_2001, davies_causal_2004, peeters_time_2005, gouyon_experimental_2006}.

With the introduction of deep learning, like most other MIR tasks, tempo estimation methods have gradually moved towards end-to-end learning where the deep neural network would jointly perform the feature extraction (e.g. onset detection curve) and tempo estimation \cite{bock_accurate_2015, schreiber_single-step_2018, bock_deconstruct_2020}.
This evolution brought a boost in performance so that the tempo estimation state of the art benchmark have been based on supervised deep learning for some years \cite{bock_accurate_2015, schreiber_single-step_2018, bock_deconstruct_2020}. 
In line with this observation, we also adopt a deep neural network model described in detail in Section \ref{sec:model}.


\subsection{Siamese networks and trivial solutions}
\label{sec:ssl}

Most of the methods that have been shown to be successful at self-supervised representation learning for domains where data is dense and continuous, such as image and audio \cite{chen_simple_2020, caron_unsupervised_2020, grill_bootstrap_2020, niizumi_byol_2021, spijkervet_contrastive_2021}, are variations of the Siamese networks architecture \cite{hadsell_dimensionality_2006}. 
Another trait these methods have in common is that they aim to maximise the similarity between the representations obtained from different transformations of a training sample.
This problem, however, admits trivial solutions. For example, if the model produces a constant output irrespective of input, the representations of the two views would always be identical. Providing a mitigation strategy to prevent the model to collapse to a trivial solution is a major challenge in this family of methods.


A variety of strategies have been proposed to guard against the trivial solution collapse issue in the literature. 
SimCLR relies on negative sampling, where for each training sample all other samples in the mini-batch are considered as negatives \cite{chen_simple_2020}. This class of methods tend to benefit from large batch sizes. 
The Barlow twins approach prevents collapse by introducing an additional redundancy term to the training loss \cite{zbontar_barlow_2021}. The authors also note that the method require smaller batch size than SimCLR.
Deep cluster \cite{caron_deep_2018} and SwAV \cite{caron_unsupervised_2020} prevents collapse by incorporating a clustering mechanism so that image features are not compared directly.
Departing from the contrastive approach, BYOL relies only on positive pairs and prevents collapse by introducing asymmetry in the training scheme by updating the model parameters using only one transformed view of the input while the other view is used as a target \cite{grill_bootstrap_2020}. In a similar line of work, the SimSiam authors note that the 'stop-gradient' operation introduces some asymmetry that is critical in preventing the collapse to trivial solutions \cite{chen_exploring_2020}. Also relying on asymmetry, MoCo mitigates the trivial solution collapse by including momentum encoder\cite{he_momentum_2020}.

The methods described above introduce trivial solution collapse mitigation strategies of varying degree of complexity that are found to work in practice even though their formulation still formally accepts trivial constant solutions. 
In Section \ref{sec:training_strategy} we propose a simple framework that formally does not admit a trivial constant solution, does not require negative samples \cite{chen_simple_2020, spijkervet_contrastive_2021}, large batches \cite{chen_simple_2020, spijkervet_contrastive_2021}, asymmetry \cite{grill_bootstrap_2020, he_momentum_2020, chen_exploring_2020, niizumi_byol_2021}, non-differentiable operators \cite{caron_unsupervised_2020} or stop gradients \cite{chen_exploring_2020}. 

\subsection{Equivariant objective}
\label{sec:equiv}

A representation $q(w)$ is invariant if it remains unchanged for a certain transformation $k$ of the input $w$:

\begin{equation}
\label{eq:invariance}
q(k \cdot w) = q(w)
\end{equation}

In the case of equivariance, the representation reflects the transformation applied to the input:

\begin{equation}
\label{eq:equivariance}
q(k \cdot w) = k \cdot q(w)
\end{equation}

Although the bulk of self-supervised methods developed so far rely on the notion of invariance to transformations of the training samples, we argue here that equivariance can provide a useful learning signal for forcing representations to capture meaningful properties of the input data. Recent works in the computer vision domain show that employing an equivariant objective in addition to more common invariant objective is beneficial \cite{dangovski_equivariant_2021, wang_self-supervised_2020}. In this scenario, the equivariance is enforced against transformations such as rotation or scale.
Similarly, equivariance is starting to be explored as an additional training objective in the video domain \cite{jenni_time-equivariant_2021}.

In the musical audio domain, SPICE relies on a composite training objective with an equivariance constraint to learn pitch representations from unlabelled data at its core \cite{gfeller_spice_2020}. SPICE generates two views of each training sample by applying pitch-shifting and aims to learn representations that are equivariant to the pitch of musical audio.
SPICE requires strong regularisation to train effectively, which is achieved by adding an extra decoder network (discarded at inference time) and corresponding audio input reconstruction term to the loss.

In contrast with previous works, the method we propose here solely relies on a simple equivariance loss term. It does not include an invariance-based objective \cite{dangovski_equivariant_2021, wang_self-supervised_2020} or any extra regularisation loss term \cite{gfeller_spice_2020}.


\section{Methods}
\label{sec:methods}

\subsection{Model}
\label{sec:model}

Our model pipeline is a close adaptation of the Temporal Convolutional Network (TCN) architecture architecture introduced first in \cite{matthewdavies_temporal_2019} for beat tracking and later extended to joint beat tracking and tempo estimation in \cite{bock_multi-task_2019}.
From the audio downmixed to mono, we compute a magnitude spectrogram with a window and FFT size of 2048 samples and a hop size of 441 samples (i.e. 100 frames per second for audio sampled at 44100Hz). The FFT magnitude spectrogram is then mapped to a Mel Spectrogram with 81 bins ranging from 30 to 17,000Hz, and finally logarithmic compression is applied.

\begin{figure}[h!]
\centering
\includegraphics[width=0.85\columnwidth]{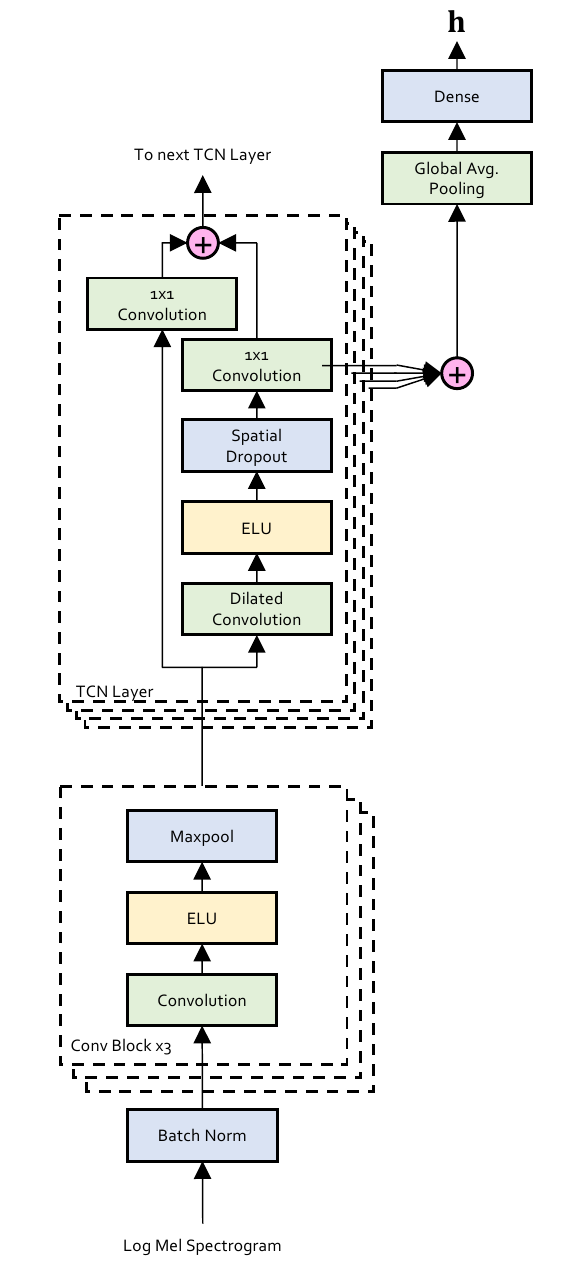}
\caption{Temporal convolutional neural network (TCN) architecture. The Log Mel spectrogram is first processed through 3 convolution blocks and then through 8 TCN layers with 16 filters and geometrically increasing dilation rates from $2^0$ to $2^7$. The network outputs a 16-d vector embedding $\mathbf{h}$.}
\label{fig:model}
\end{figure}

The Log Mel Spectrogram is then fed to a neural network constituted of two main building blocks and which architecture is depicted in Figure \ref{fig:model}. The input is first fed through a batch normalisation layer followed by 3 convolutional blocks. It is then processed through 8 TCN layers with 16 filters each and geometrically increasing dilation rates from $2^0$ to $2^7$. 
Because we focus on learning tempo representations in this work, we drop the beat tracking branch and adapt the tempo branch architecture so that our network outputs a 16-d vector representation $\mathbf{h}$. All other design parameters are identical to \cite{bock_multi-task_2019}. 
We chose this TCN architecture because it is specifically designed to model temporal characteristics of music and has been shown to provide very competitive performance despite using very little parameters (33k).

\subsection{Training strategy}
\label{sec:training_strategy}

The objective of the training strategy described below is to learn representations that capture tempo information without having access to tempo annotations at training time. Starting from the observation that the time-stretching modifies the tempo of a music piece, we propose to use equivariance to time-stretching as a self-supervision objective to learn tempo representations.


\begin{figure}[h]
\label{fig:sst_diagram}
\centering
\includegraphics[width=0.8\columnwidth]{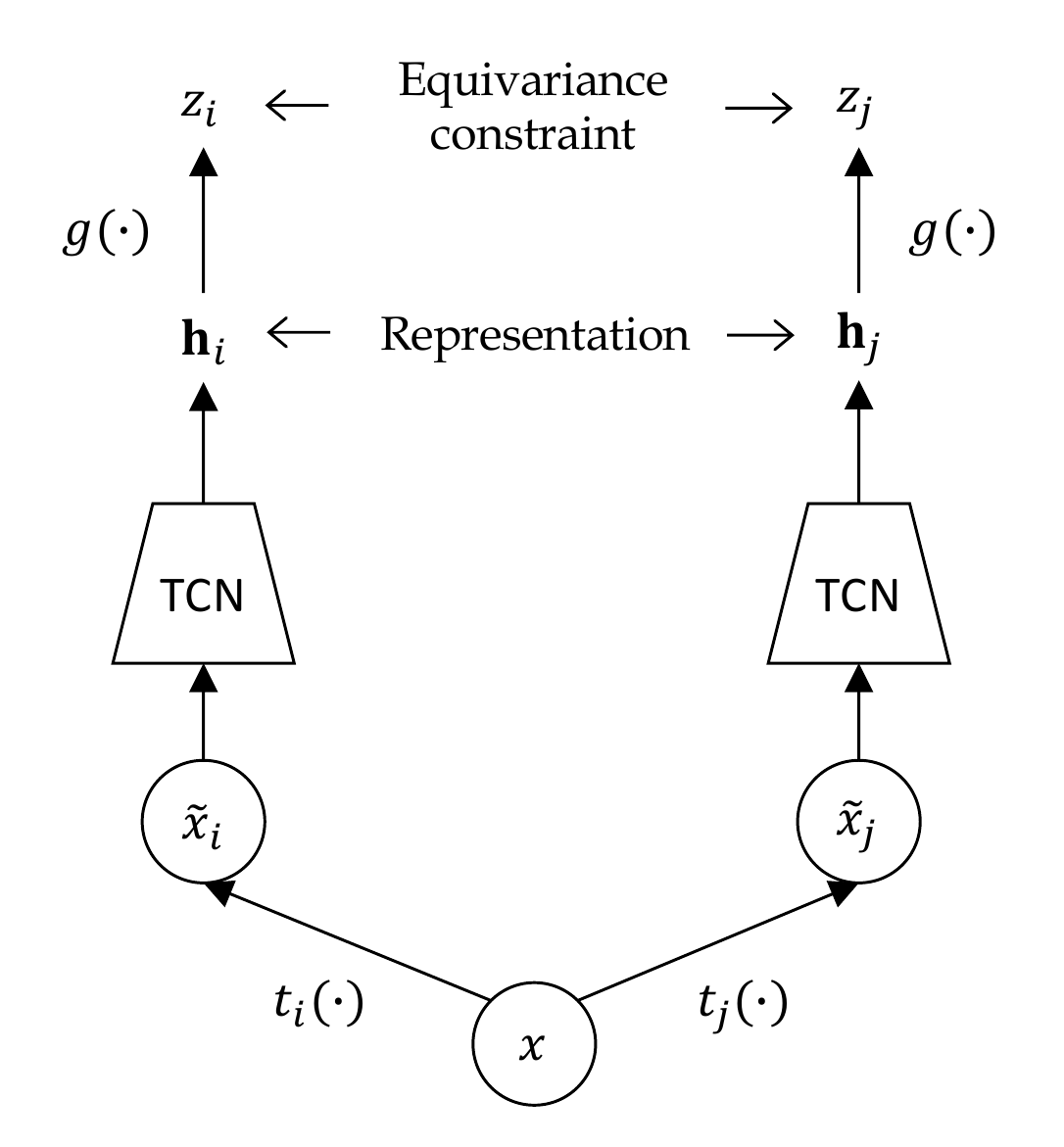}
\caption{Equivariant self-supervision framework. Two distinct time-stretching transformations ($t_i$ and $t_j$) are applied to a training sample $x$ to obtain to correlated views ($\tilde{x}_i$ and $\tilde{x}_j$). The TCN network and projection head $g(\cdot)$ are trained to produce a pseudo-tempo scalar $z$ that is equivariant to the time stretching transformation of the input.
The projection head is discarded after training.}
\end{figure}

For a recording taken from the training set, let $x$ be an excerpt of length $l_x$ selected at random in the recording, with unknown tempo $y \in \mathbb{R}$. 
For each sample $x$, two views $(\tilde{x}_i, y_i)$ and $(\tilde{x}_j, y_j)$ are produced by applying time stretching transformations noted $t_i$ and $t_j$ respectively using Sox\footnote{http://sox.sourceforge.net}. For each view, the time stretching rate $\alpha$ is drawn uniformly at random from $[1-r, 1+r]$ where $r$ is a hyper-parameter to the training procedure. As a result, we obtain:

\begin{align}
\label{eq:tempo_aug_1}
\tilde{x}_i &= t_i(x), & y_i &= \alpha_i \cdot y \\
\label{eq:tempo_aug_2}
\tilde{x}_j &= t_j(x), & y_j &= \alpha_j \cdot y 
\end{align}

where the right hand side of eq. (\ref{eq:tempo_aug_1}) and eq. (\ref{eq:tempo_aug_2}) materialises the transformation of the tempo of each view. 

In order to allow efficient batch processing at training time, we force the augmented views $\tilde{x}_i$ and $\tilde{x}_j$ to all have the same length $l_x$ by cropping if they are longer and padding with zeros if shorter. In all our experiments we set $l_x$ to be 600,000 audio samples (13.6s at 44.1 kHz).

Because the true tempo $y$ is unknown, we propose to use a pseudo-tempo representation $z \in \mathbb{R}$ as the output of model at training time. 
Let $f(\cdot)$ be the transformation applied by the TCN and $g(\cdot)$ be a linear projection head, so that $g(f(x)) = z$. 
The objective is then to constrain the TCN and projection head stack to be equivariant to the time-stretching transformation of the input, so that: 

\begin{equation}
\label{eq:tcn_equivariance}
g\left(f\left(t(x)\right) \right)= \alpha \cdot z
\end{equation}

Since the two views are derived from the same training sample, it is trivial to show that the equivariance formulation expressed in eq. (\ref{eq:tcn_equivariance}) yields: 

\begin{equation}
\label{eq:tcn_equivariance_2}
\alpha_i \cdot z_j = \alpha_j \cdot z_i
\end{equation}

In other words, the equivariance objective is met if eq. (\ref{eq:tcn_equivariance_2}) is true. 
Based on this, we can derive the following training loss that is minimised when the equivariance objective is met: 

\begin{equation}
\label{eq:loss}
\mathcal{L} = \left| \frac{z_i}{z_j} - \frac{\alpha_i}{\alpha_j} \right|
\end{equation}


Note how this formulation does not allow the model to collapse on a trivial constant solution to minimise the loss. Producing a constant $z$ value for any input does not yield a minimal loss because $\alpha$ values are drawn at random for every training sample, which means that the ratio $\frac{\alpha_i}{\alpha_j}$ varies for every pair of training sample views.

Other formulations of the loss function that are minimised when eq. (\ref{eq:tcn_equivariance_2}) is true can be derived, but may allow trivial solutions. For example loss functions such as $\mathcal{L}' = \left| \alpha_i \cdot z_j - \alpha_j \cdot z_i \right|$ or $\mathcal{L}'' = \left| z_i -  \frac{\alpha_i \cdot z_j}{\alpha_j} \right|$ admit a trivial optimal solution for $z_i = z_j = 0$.

\subsection{Training parameters}

In all our experiments we use a batch size of 16 samples, the Adam optimiser \cite{kingma_adam_2014} with initial learning rate of 0.001. We pre-train the model for 20 epochs and fine-tune for 100 epochs. 

With the aim to promote robustness against non tempo-related attributes of audio signals, we add the following optional random audio augmentations during pre-training: gain, polarity inversion, gaussian noise and SpecAugment frequency masking \cite{park_specaugment:_2019}. Note that our loss function remains unchanged whether or not we apply these augmentations.

\subsection{Datasets}

For self-supervised training we use the MagnaTagaTune dataset (MTT) \cite{law_evaluation_2009}. It contains around 25k audio tracks but no tempo annotations. 

For fine-tuning and evaluation we use datasets commonly used in the tempo estimation litterature that do contain tempo annotations. 
Namely we use the following datasets: GTZAN \cite{tzanetakis_musical_2002}, Hainsworth \cite{hainsworth_particle_2004} , Giantsteps \cite{knees_two_2015, schreiber_crowdsourced_2018} and ACM Mirum \cite{peeters_perceptual_2012}. 

\begin{figure*}[t]
\centering
\includegraphics[width=1.6\columnwidth]{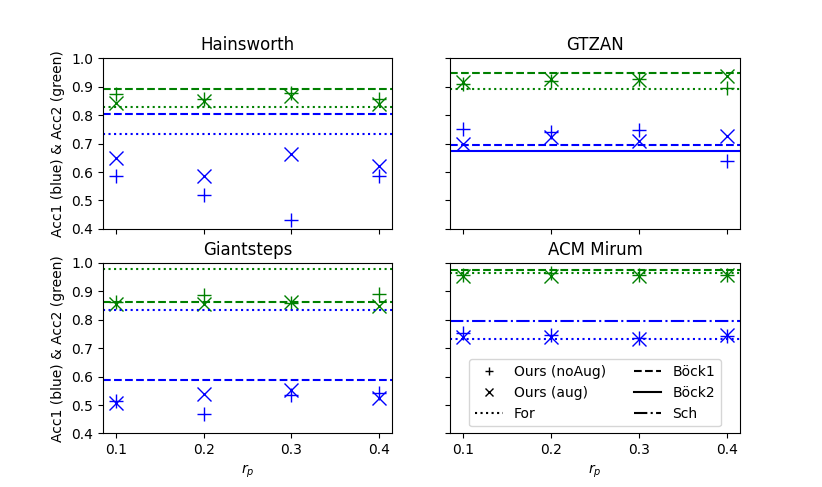}
\caption{Performance metrics for our proposed method, compared against supervised benchmarks. We report both Accuracy 1 (in blue) and Accuracy 2 (in green). We report results for 8 pre-training conditions, which are combinations of using  augmentations ("aug") or not ("noAug") and of $r_p \in \{0.1, 0.2, 0.3, 0.4\}$. In all cases the model is fine-tuned with a time-stretching of strength $r_f=0.2$.
Horizontal lines represent the supervised benchmarks "Böck1" \cite{bock_accurate_2015}, "Böck2" \cite{bock_multi-task_2019}, "For" \cite{foroughmand_deep-rhythm_2019}, "Sch" \cite{schreiber_single-step_2018}. In the interest of legibility, for every dataset and every metric we only display the highest and lowest performing baselines. 
}
\label{fig:acc_sst}
\end{figure*}

\subsection{Evaluation}

After pre-training, we wish to evaluate the representations learnt by the TCN. The projection head $g(\cdot)$ is discarded and all the network weights frozen. We then attach a linear classification head with 300 units and softmax layer to represent the range [0,300] BPM, apply a smoothing window to the ground truth label similar to \cite{bock_multi-task_2019} and fine-tune using the cross-entropy loss. 

In order to evaluate generalisation capability of our model, we perform cross-dataset evaluation. 
This means that the metrics we report are always computed on a dataset that has never been seen by the model during the pre-training or fine-tuning stages.

In order for our results to be comparable with existing literature, we report tempo performance metrics \emph{Accuracy 1} and \emph{Accuracy 2} scores with a $\pm 4\%$ tolerance \cite{gouyon_experimental_2006}. 
\emph{Accuracy 1} measures the accuracy of the model's prediction of the exact ground truth tempo, while \emph{Accuracy 2} also allows for "octave errors" with factor $\{2, 3, \frac{1}{2}, \frac{1}{3} \}$. We leave further evaluation considerations for future work \cite{schreiber_music_2020}.

\section{Experiments}
\label{sec:experiments}

\subsection{Robustness against trivial solutions}

In a preliminary experiment, we ran the pre-training with alternative losses $\mathcal{L}'$ and $\mathcal{L}''$, as defined in Section \ref{sec:training_strategy}. It systematically collapsed to a trivial solution $z\approx0$. Conversely, we use the loss function $\mathcal{L}$ described in Eq. \ref{eq:loss} in all the experiments reported in this paper and did not observe collapse, which confirms its robustness against trivial constant solutions.

\subsection{Influence of pre-training augmentation parameters}
\label{sec:aug_param_ssl}

Figure \ref{fig:acc_sst} shows the performance metrics on evaluation datasets for supervised baselines and our method after pre-training on the MTT dataset and cross-dataset fine-tuning and evaluation.
We report results for 8 pre-training conditions, which are combinations of using audio augmentations ("aug") or not ("noAug"), and of the strength of time-stretching during pre-training $r_p \in \{0.1, 0.2, 0.3, 0.4\}$. In all cases the model is fine-tuned with a time-stretching of strength $r_f=0.2$ (see section \ref{sec:finetune_aug}).

It appears that adding augmentations yields no notable performance benefit except on Acc1 on the Hainsworth dataset. We hypothesise this is because the TCN architecture already has a strong inductive bias towards temporal and rhythmic structures, which makes it robust against potentially confounding attributes of the audio signal. 

The choice and strength of input transformations has been shown to have an impact on the performance of contrastive learning \cite{tian_what_2020}. Intuitively, one expects that very gentle augmentations may not allow to learn robust representations, on the other hand too strong an augmentation may lose semantic content and therefore not allow to learn at all.
In our experiments, we observe that the performance tends to dip slightly at the extreme ends of the range of values for $r_p$. However, it is interesting to note that we do not observe a dramatic degradation of performance even though the transformations applied then (e.g. $r_p=0.4$) are musically extreme.  

Overall these results suggest that the pre-training phase is fairly robust to the choice of time-stretching parameters and to the absence of audio augmentation.

\subsection{Influence of fine-tuning augmentation parameters}
\label{sec:finetune_aug}

\begin{table*}[t]
\centering
\begin{tabular}{c c c c c c c c c c} 
 \toprule
 \multirow{2}{*}{Method} & \multirow{2}{*}{$r_f$} & \multicolumn{2}{c}{Hainsworth} & \multicolumn{2}{c}{GTZAN} & \multicolumn{2}{c}{Giantsteps} & \multicolumn{2}{c}{ACM Mirum} \\ 
  & & Acc1 & Acc2 & Acc1 & Acc2 & Acc1 & Acc2 & Acc1 & Acc2 \\ [0.5ex] 
 \hline
Ours & 0.0   & 0.604 & \textit{0.838} & \textit{0.691} & 0.887 & 0.512 & 0.809 & 0.704 & 0.943 \\
Ours & 0.1   & 0.586 & 0.824 & \textit{0.719} & 0.881 & 0.456 & 0.791 & \textit{0.757} & \textit{0.965} \\
Ours & 0.2   & 0.518 & \textit{0.856} & \textit{0.741} & \textit{0.919} & 0.470 & \textit{0.886} & \textit{0.747} & \textit{0.965} \\ 
Ours & 0.3   & 0.550 & \textit{0.829} & \textbf{0.785} & \textit{0.921} & 0.438 & 0.846 & 0.700 & 0.958 \\ 
Ours & 0.4   & 0.541 & \textit{0.829} & \textit{0.778} & \textit{0.926} & 0.472 & \textit{0.884} & 0.724 & 0.952 \\ 
\hline
Schreiber \cite{schreiber_single-step_2018}  &  -   & 0.770 & 0.842 & 0.694 & 0.926 & 0.730 & 0.893 & \textbf{0.795} & 0.974 \\
Foroughmand \cite{foroughmand_deep-rhythm_2019} & - & 0.734 & 0.829 & 0.697 & 0.891 & \textbf{0.836} & \textbf{0.979} & 0.733 & 0.965 \\
Böck 1 \cite{bock_accurate_2015}  &  -        & \textbf{0.806}& \textbf{0.892} & 0.697 & \textbf{0.950} & 0.589 & 0.864 & 0.741 & \textbf{0.976} \\
Böck 2 \cite{bock_multi-task_2019}  &  -              &   -   &   -   & 0.673 & 0.938 & 0.764 & 0.958 & 0.749 & 0.974 \\ [1ex] 
 \bottomrule
 \end{tabular}
\caption{Influence of time stretching parameters during Fine-tuning, and comparison to supervised baselines. In all cases, our model is pre-trained on MTT with $r_p=0.2$ and no audio augmentations. Highest performance for each metric and dataset shown in bold. Metrics where our model outperforms at least one of the baselines in italics.}
\label{table:finetune}
\end{table*}

At the fine-tuning stage, tempo estimation is formulated as a multiclass classification problem, where each class corresponds to a tempo value. Because it is likely that the datasets used for fine-tuning may not contain training samples for each tempo in the full BPM range considered, we also apply a time-stretching augmentation to increase the range of tempi seen during fine tuning. 

Table \ref{table:finetune} summarises the results using a model pre-trained with $r_p=0.2$ and no invariant augmentations, for a range of fine-tuning time-streching augmentation strength $r_f$. 
As expected, it appears that applying some time-stretching ($r_f>0$) generally improves performance over not applying any ($r_f=0$). 
We also note that the optimal value varies from one dataset to the next. For example results seem to be optimal for relatively large augmentation strength on the GTZAN dataset while they would be optimal for smaller values on ACM Mirum.

\subsection{Comparison to supervised benchmarks}

Figure \ref{fig:acc_sst} shows supervised benchmarks as horizontal lines. 
Our proposed method's Accuracy 2 performance is comparable with supervised benchmarks on all datasets. 
Accuracy 1 performance is more inconsistent. It lags behind supervised methods on Hainsworth and Giantsteps datasets, while it is comparable with supervised benchmarks on ACM Mirum. Notably, our method outperforms all supervised baselines in Accuracy 1 on GTZAN.
Similar conclusions can be drawn from the results shown in Table \ref{table:finetune}, under different fine-tuning configurations. 

Note that we evaluate our models on datasets that have never been seen during pre-training or fine-tuning and still observe performance generally competitive with supervised benchmarks. 
Also taking into account that we only fine-tune a linear layer, this result indicates a promising degree of robustness of the representation learnt during pre-training against domain shift.



\section{Closing words}
\label{sec:conclusion}


In this work, we introduced an approach to use equivariance as a self-supervision signal to learn audio tempo representations from unlabelled data. We derive a simple loss function that prevents the network from collapsing on a trivial solution during training, without requiring any form of regularisation or negative sampling.
Our experiments show not only that it is possible to learn meaningful representations for tempo estimation by solely relying on equivariant self-supervision, but also demonstrate that we can achieve performance comparable with supervised methods on most benchmarks. 
We also show that the representations learnt exhibit promising robustness against pre-training and fine-tuning hyper-parameters as well as against domain shift. 
As an added benefit, our method only requires moderate compute resources by making use of a small model and not requiring large batch sizes, therefore keeping it accessible to a wide research community.

We believe these results open a number of interesting avenues for future work. 
This paper is focused on tempo estimation but there is potential to extend our investigation to other MIR tasks revolving around rhythmic properties such as beat tracking, metrical structure estimation or rhythm pattern identification.
In addition, because no annotated data is required for pre-training, there is potential for investigating applications to low resource musical genres or genres underserved by traditional and current supervised methods.
Since our loss function is very simple, it could also be added as an extra objective in SSL frameworks for learning general music representations, at a moderate cost of increased complexity. 
Last but not least, while invariance typically used in contrastive learning is naturally connected with classification problems, equivariant self-supervision is well suited to tasks that can be formulated as regression problems. We therefore believe there is potential to explore many more applications of equivariant SSL in the music domain and beyond.

\bibliography{bibliography}

%
%
%
%
%

\end{document}